\documentclass[12pt,reqno]{amsart}
\title[\protect\parbox{0.95\textwidth}{Bounds on the Global Attractor of 2D Incompressible Turbulence in
the Palinstrophy--Enstrophy--Energy Space}]{Bounds on the Global Attractor of 2D Incompressible
Turbulence in the Palinstrophy--Enstrophy--Energy Space}
\author{Pedram Emami}
\author{John C. Bowman}
\address{Department of Mathematical and Statistical Sciences, University of Alberta, Edmonton, Alberta
T6G 2G1, Canada}

\usepackage{bm}
\usepackage{float}
\usepackage{graphicx}
\usepackage{amsmath,amssymb,amsthm,amsfonts}
\usepackage{lineno}
\usepackage{mathdef}
\usepackage{hyperref}
\usepackage{url}

\usepackage[style=ext-authoryear,citestyle=ext-authoryear,hyperref=true,natbib=true]{biblatex}
\DeclareOuterCiteDelims{parencite}{\bibopenbracket}{\bibclosebracket}
\DeclareInnerCiteDelims{textcite}{\bibopenbracket}{\bibclosebracket}
\let\cite\parencite
\let\citename\textcite

\def\asygl{https://www.math.ualberta.ca/~bowman/asygl/bilinear}

\def\includeFigure#1#2{
\begin{center}
\includegraphics{fig/#1}
\caption{#2}\label{#1}
\end{center}
}

\def\includeWebFigure#1#2{
\begin{center}
\href{\asygl/#1}{\includegraphics{fig/#1}}
\caption{#2 Click on the figure to view a 3D interactive image.}\label{#1}
\end{center}
}

\newtheorem*{definition}{Definition}
\newtheorem{theorem}{Theorem}

\newtheorem*{remark}{Remark}
\newtheorem*{prf}{Proof}

\def\v{\bm}

\def\clap#1{\hbox to 0pt{\hss#1\hss}}

\def\endofprf{\hfill$\blacksquare$}

\makeatletter
\let\cal\mathcal

\def\hyperindex#1{\index{#1}\hypertarget{#1}}
\def\In#1{\hyperindex{#1}{\emph{#1}}}

\def\grad{\v\nabla}
\def\div{\grad\dot}
\def\lap{{\nabla}^2}
\def\lapinv{{\nabla}^{-2}}
\def\cross{{\v\times}}
\def\vdot{{\v \cdot}}
\def\etal{{\it et al.}}

\def\clap#1{\hbox to 0pt{\hss#1\hss}}
\def\endofprf{\hfill$\blacksquare$}

\def\hyperindex#1{\index{#1}\hypertarget{#1}}
\def\Eq#1{(\ref{#1})}

\def\ddt{\frac{d}{dt}}
\def\inv{^{-1}}
\def\hf{\frac{1}{2}}
\def\bhf{\bgroup1/2\egroup}
\def\Ahf{\bgroup A^{1/2}\egroup}
\def\textoversign#1#2{\overset{\text{#1}}{#2}}
\def\ubar#1{\underline{\sbox\tw@{$#1$}\dp\tw@\z@\box\tw@}}
\def\N{\mathbb{N}}
\def\Z{\mathbb{Z}}

\def\R{\mathbb{R}}
\def\B{{\cal B}}

\def\G{\tilde G}
\def\|#1|{\@ifnextchar|{\left\lVert#1\right\rVert\sbox0}{\left|#1\right|}}
\makeatother

\definecolor{heavyred}{cmyk}{0,1,1,0.25}
\definecolor{heavyblue}{cmyk}{1,1,0,0.25}
\hypersetup{
  pdftitle={},
  pdfpagemode=UseOutlines,
  citebordercolor=0 0 1,
  colorlinks=true,
  allcolors=heavyred,
  breaklinks=true,
  pdfauthor={Pedram EmamiC and John C. Bowman},
  pdfpagetransition=Dissolve,
}

\newtheoremstyle{roStyle}{}{}{}{}{\it\bfseries}{.}{ }{}
\newtheoremstyle{exStyle}{}{}{}{}{\bfseries}{.}{ }{}  

{
  \theoremstyle{roStyle}
  \newtheorem{observation}{Observation}
 
}
{
  \theoremstyle{exStyle}
  
}

\bibliography{refs, refsOT}
\begin{document}
\maketitle
\begin{abstract}
Analytic bounds on the projection of the global attractor of 2D incompressible turbulence in the
palinstrophy--enstrophy plane [Dascaliuc, Foias, and Jolly 2005, 2010] are observed to vastly
overestimate the values obtained from numerical simulations. This is due to the lack of a good estimate
for the inner product $(\cal{B}(\vu,\vu),A^2\vu)$ of the advection term and the biLaplacian. Sobolev
inequalities like Ladyzhenskaya or Agmon's inequalities yield an upper bound that we show is not sharp.
In fact, for statistically isotropic turbulence, the expected value of $(\cal{B}(\vu,\vu),A^2\vu)$ is
zero. The implications for estimates on the behaviour of the global attractor are discussed.
\end{abstract}
\section{Introduction}
One of the most recent approaches in studying turbulence incorporates functional analysis tools to study
turbulence on a solid mathematical basis. The Navier--Stokes equation is the generally accepted governing
partial differential equation of turbulence. Unlike empirical or heuristic formulations, it provides a
solid foundation for studying important intrinsic characteristics of turbulence.

Building on the work of \citename{Dascaliuc2010} that characterized the bounds on the global attractor of
2D homogeneous incompressible turbulence in the energy--enstrophy and enstrophy--palinstrophy planes for
constant forcing, we extended in \cite{Emami18} the analysis to handle random forcing, which is more
applicable to both laboratory and numerical experiments. For white-noise random forcing, corresponding
bounds on the global attractor were obtained  on the energy--enstrophy plane and demonstrated with direct
numerical simulations. While it turned out that the same form of the upper bound found for constant
forcing also holds for random forcing, it was observed that the projected upper bound of the global
attractor on the enstrophy--palinstrophy plane vastly overestimates numerically computed trajectories.

These observations motivated us to study the upper bound of the projection of the global attractor onto
the enstrophy--palinstrophy plane in more detail. The result is a much tighter upper bound that, together
with the upper bound on the energy--enstrophy plane, provides a better understanding of the global
attractor as well as a more discerning criteria for validating heuristic engineering subgrid models.

\section{Definitions and preliminaries}
One of the simplest contexts in which to pose the turbulence problem is 2D incompressible
homogeneous isotropic turbulent flow in a bounded domain with periodic boundary conditions and zero
mean velocity and forcing. One close realization of this ideal form of turbulence in laboratories
is a very thin layer of turbulent fluid far downstream from a flow passing over a net of wires.

Looking at this ideal form of turbulence deterministically involves using the incompressible
Navier--Stokes and continuity equations expressed as a set of integro-differential equations, with
zero mean flow and forcing and constant density $\rho=1$:
\begin{align}\label{2D-governing-equations}
&\dfrac{\partial\vu}{\partial t}-\nu{\nabla}^2\vu+\vu\dot\grad\vu+\grad p=\vF,\\
&\div \vu = 0,\quad\int_{\Omega}{\vu\,d\vx}=\bm{0},\quad\int_{\Omega}{\vF\,d\vx}=\bm{0},\\
&\vu(\vx,0)=\vu_0(\vx),
\end{align}
with $\Omega=\[0,L\]\times \[0,L\]$ and periodic boundary conditions on $\partial\Omega$. This
problem can be considered in a specific Hilbert space $H$ with an
extended $L^2$ inner product that captures the random nature of the
applied forcing:
\begin{equation*}
\(\vu,\vv\)\doteq\int_{\Omega}{\<\vu\dot\vv\>\,d\vx}=\int_{\Omega}{
\(\int_{-\infty}^\infty{\vu\dot\vv\,\dfrac{dP}{d\zeta}d\zeta}\)\,d\vx},
\end{equation*}
where $P$ is a distribution function. For each realization $u$, the Hilbert space is defined as
\begin{equation}\label{Hilbert-space-def}
H(\Omega) \doteq \cl\left\{\vu \in (C^2(\Omega)\intersect L^2(\Omega))^2\ |\ \div
\vu=0,\ \int_{\Omega}{\vu\, d\vx = 0}\right\},
\end{equation}
with the extended $L^2$ norm
\begin{equation*}
\|\vu||=(\vu,\vu)^{1/2}=\(\int_{\Omega}\int_{-\infty}^{\infty}{\vu(\vx,t)\dot\vu(\vx,t)
\dfrac{dP}{d\zeta}\,d\zeta d\vx}\)^{1/2}.
\end{equation*}
From now on, we will drop the term `extended' from both inner product and the resulting norm since
that does not cause any confusion. Here, $\doteq$ is used to emphasize a definition and
$\cl$ denotes the closure with respect to the $L^2$ norm.

The above problem can then be expressed as
\begin{equation}
\dfrac{d\vu}{dt} - \nu\lap\vu + \vu\dot\grad\vu + \grad p =
\vF,\qquad \vu(t)\in H(\Omega).
\end{equation}
Let $A\doteq-\cal P(\lap)$, $\vf\doteq\cal P(\vF)$, and define the bilinear map
\begin{equation*}
\cal B(\vu,\vu)\doteq\cal P \left(\vu\dot\grad\vu + \grad p\right),
\end{equation*}
where $\cal P$ is the Helmholtz--Leray projection operator
from $(L^2(\Omega))^2$ to $H(\Omega)$:
\begin{equation*}
\cal P(\vv)\doteq\vv - \grad\lapinv\div\vv, \qquad \forall \vv \in
(L^2(\Omega))^2.
\end{equation*}
In terms of these definitions, \Eq{Hilbert-space-def} can be written more
compactly as
\begin{equation}
\dfrac{d\vu}{dt} + \nu A\vu + \cal{B} (\vu,\vu) = \vf.\label{NSFoias}
\end{equation}
\section{Stokes operator \texorpdfstring{$A$}{A}}
The operator $A=\cal P(-\lap)$ is positive-semidefinite and self-adjoint
in~$H(\Omega)$, with a compact inverse and eigenvalues
\begin{equation*}
\lambda=k_0^2\vk\dot\vk, \qquad \vk \in
\Z\times\Z\backslash\{\bm{0}\},
\end{equation*}
where $k_0=2\pi/L$.
The eigenvalues of this positive-definite infinite-dimensional linear operator can
be arranged as
\begin{equation*}
0<\lambda_0<\lambda_1<\lambda_2< \cdots,\qquad \lambda_0=\(\dfrac{2\pi}{L}\)^2.
\end{equation*}
The corresponding eigenvectors $\v w_i,\ i\in \N_0$ form an orthonormal basis for the
Hilbert space $H$ upon which we can define any power of $A$:
\begin{equation*}
A^\alpha \v w_j=\lambda_j^\alpha \v w_j, \qquad \alpha \in \R,\quad j \in \N_0.
\end{equation*}
Having the above orthonormal basis, we can use our inner product
to define new spaces $V^{2\alpha}~\subset~H$
for any real value of $\alpha$: \cite{Temam95Navier}
\begin{equation*}
V^{2\alpha}=D(A^{\alpha})\doteq\left\{\vu \in H\ |\ \sum_{j=0}^\infty\lambda_j^{2\alpha} (\vu,\v w_j)^2
<\infty\right\}.
\end{equation*}

Among all possible values of $\alpha$, we are especially interested in the spaces $V^0, V^1$, $V^2$, and
$V^3$ corresponding to $\alpha=0, \alpha=1/2$, $\alpha=1$, and $\alpha=3/2$ motivated by the definition
of energy, enstrophy, palinstrophy, and hyperpalinstrophy. It is readily shown that
\begin{align*}
\vu\in V^0	 &\iff \|\vu|| <\infty,\\
\vu\in V^1   &\iff \|A^\bhf\vu|| <\infty,\\
\vu\in V^2	 &\iff \|A\vu|| <\infty,\\
\vu\in V^3	 &\iff \|A^{3/2}\vu|| <\infty.
\end{align*}

For any solution $\vu$ of the 2D Navier--Stokes equation, we
define the $\zeta$th-order \emph{polystrophy}
$$
E_\zeta(t) = \half\|A^{\zeta/2}\vu(t)||^2.
$$

The special polystrophy values $E_0$, $Z\doteq E_1$, $P\doteq E_2$,
$P_2\doteq E_3$ are respectively known as the \emph{energy},
\emph{enstrophy}, \emph{palinstrophy}, and
hyperpalinstrophy \cite{rondoni1999,al2012}.

Just as energy is proportional to the mean-squared velocity, enstrophy is proportional to the
mean-squared vorticity and therefore provides a measure of the
rotational energy in a flow. It is easily shown that the rate at
which energy is dissipated is proportional to the enstrophy. Likewise,
enstrophy is dissipated at a rate proportional to the palinstrophy.

\section{The Navier--Stokes equations with random forcing as a dynamical system for energy,
enstrophy, and palinstrophy}

On taking the inner product of $\vu$, $A\vu$, and $A^2\vu$
respectively with \Eq{NSFoias}, one obtains the following dynamical system equations for energy, enstrophy, and palinstrophy:
\begin{align}
&\hf\ddt\|\vu(t)||^2+\nu\|A^\bhf\vu(t)||^2+(\cal B(\vu,\vu),\vu)=(\vf,\vu(t)),
\label{NS-dim-energy}   \\
&\hf\ddt\|A^\bhf\vu(t)||^2+\nu\|A\vu(t)||^2+(\cal B(\vu,\vu),A\vu)=(\vf,A\vu(t)),
\label{NS-dim-enstrophy}\\
&\hf\ddt\|A\vu(t)||^2+\nu\|A^{3/2}\vu(t)||^2+(\cal B(\vu,\vu),A^2\vu)=(\vf,A^2\vu(t)).
\label{NS-dim-palinstrophy}
\end{align}
To study these equations further, it is necessary
to understand the triplet terms
$(\cal B(\vu,\vu),A^\zeta\vu)$ for $\zeta\in\{0,1,2\}$. As we will see later, it turns out that these
triplet terms have crucial roles in the behaviour of the global attractor of 2D turbulence.

To obtain estimates of the triplet terms, it is essential to exploit
properties of the bilinear map ${\cal B}$ along with incompressibility, periodicity, and the
self-adjointness of the Stokes operator~$A$. Here we only list the
identities for triplet terms; the reader who is interested in their
proofs is referred to the Appendix~A of \cite{Emami18}:
\begin{align}
&(\cal B(\vu,\vv),\v w)=-(\cal B(\vu,\v w),\vv),&\text{antisymmetry}\label{antisymmetry}\\
&(\cal B(\vu,\vu),A\vu)=0,				 		&\text{orthogonality in 2D}\label{orthogonality}\\
&(\cal B(A\vv,\vv),\vu)=(\cal B(\vu,\vv),A\vv), &\text{strong form of invariance in 2D}
\label{2D-strong}\\
&(\cal B(A\vu,\vu),\vu)+(\cal B(\vv,A\vv),\vu)\notag\\
&\phantom{(\cal B(A\vu,\vu),\vu)}+(\cal B(\vv,\vv),A\vv)=0. &\text{2D general identity}
\label{2D-general}
\end{align}
While \Eq{antisymmetry} and \Eq{orthogonality} clearly imply that the triplet terms in
\Eq{NS-dim-energy} and \Eq{NS-dim-enstrophy} vanish, the same cannot be shown for $(\cal
B(\vu,\vu),A^2\vu)$. However, equations \Eq{2D-strong} and \Eq{2D-general} together with Agmon's
inequality yield an upper bound found by \citename{foias2002statistical}, a more
physically-oriented proof of which is given in \citet[Appendix~A]{Emami18}.

\begin{figure}[htpb]
\begin{center}
\includegraphics{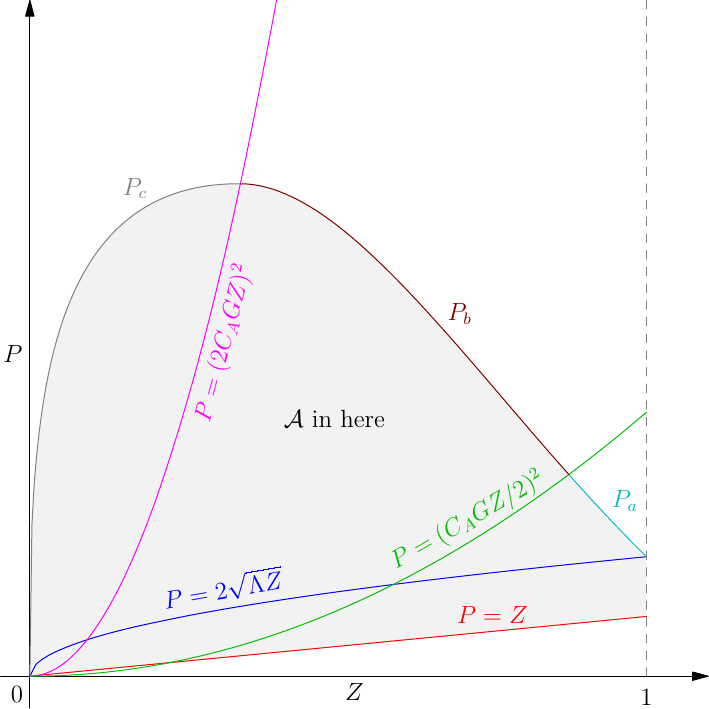}
\caption{Bounds in the normalized $Z$--$P$ plane, where
$G\doteq\protect\|\vf||/(\nu^2 k_0^2)$ and $C_A$ is the constant in
Agmon's inequality~\protect\cite{Dascaliuc2010}.}\label{PZBounds}
\end{center}
\end{figure}

Using this estimation together with some extra reasonable conditions, Dascaliuc \etal\
\cite{Dascaliuc2010} found the bounds shown schematically in
Fig.~\ref{PZBounds} for the projection of the global attractor on the
enstrophy--palinstrophy plane.

\begin{figure}[H]
\begin{center}
\includegraphics{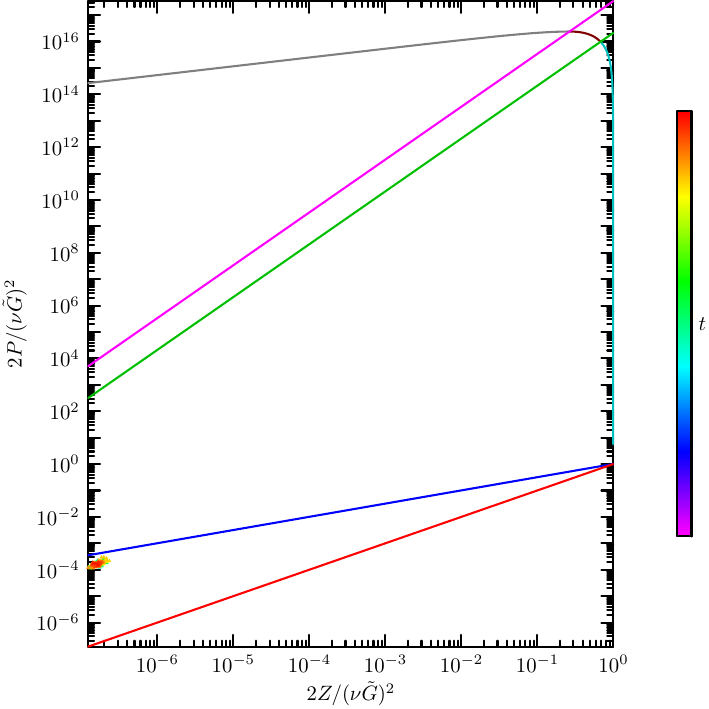}
\caption{Projection of a normalized numerical solution of 2D
turbulence with large-scale friction onto the
$Z$--$P$ plane, where $\G\doteq\sqrt{\nu\epsilon k_0^2 + \epsilon\gamma}/(\nu^2 k_0^2)$.}\label{pvzz1023}
\end{center}
\end{figure}

Unfortunately, the piecewise upper bound found by Dascaliuc \etal\
consisting of the curves $P_a$, $P_b$, and $P_c$ in
Fig.~\ref{PZBounds} is not tight enough to be provide any useful
information about the global attractor of actual turbulent
flow. Indeed, as seen in Fig.~\ref{pvzz1023}, it vastly
overestimates the projection of the global attractor for the direct
numerical solution of 2D incompressible homogeneous isotropic
turbulence with random forcing (and large-scale friction) described
in Section~\ref{pseudospectral}.

Moreover, the estimated upper bound
obtained by applying Agmon's inequality and the Cauchy--Schwartz
inequality cannot be sharp even in 3D turbulence.
For this bound to be sharp: $\B(\vu,\vu)=\a A^n\vu$ a.e.\ for some $\a\in\R$.
From the self-adjointness of $A$, such an alignment would require
\begin{align*}
0=(\B(\vu,\vu),\vu)&=(\a\vA^n\vu,\vu)=(\a\vA^{n/2}\vu,\vA^{n/2}\vu)
=\a\|A^{n/2} \vu||^2\quad\implies\\
\B(\vu,\vu)&=0 \text{ a.e.},
\end{align*}
which would imply no cascade!

In all of our numerical simulations of randomly forced isotropic
turbulence, we noticed that the triplet term is negligible.
The robustness of this observation motivated us to prove the following
result.

\begin{theorem}\label{triplet-vanishes}
For incompressible homogeneous statistically isotropic 2D turbulence
\begin{equation*}
(\cal B(\vu,\vu),A^\zeta\vu)=0\qquad\forall\zeta\in\R.
\end{equation*}
\end{theorem}
\begin{prf}
Since for 2D turbulence, the stream function $\psi(x,y)$ is well
defined, taking into account that $A$ is self-adjoint, we know that
\begin{align*}
u=\psi_y, v=-\psi_x\quad
(\vu\cdot\grad)\vu =
\begin{bmatrix}
\psi_y\psi_{yx}-\psi_{x}\psi_{yy}\\
\psi_x\psi_{xy}-\psi_{y}\psi_{xx}\\
\end{bmatrix}
,\quad A^\zeta\vu =
\begin{bmatrix}
A^\zeta u\\
A^\zeta v
\end{bmatrix}.
\end{align*}
Therefore,
\begin{align*}
(\vu\cdot\grad)\vu\cdot A^\zeta\vu &=
A^\zeta\psi_y(\psi_y\psi_{yx}-\psi_{x}\psi_{yy})-A^\zeta\psi_x(\psi_x\psi_{xy}-\psi_{y}\psi_{xx}).
\end{align*}
Since $A$ is an isotropic operator, the ensemble average of the above
expression will vanish for isotropic turbulence. That is,
\begin{align*}
\left\langle(\vu\dot\grad)\vu\dot A^\zeta\vu\right\rangle = 0;
\end{align*}
this implies that
\begin{align*}
\(\cal B(\vu,\vu),A^\zeta\vu\)
&=\int_\Omega \<\(\vu\cdot\grad\vu\)\cdot A^\zeta\vu\>\,d\vx
+ \int_\Omega \<\grad p\cdot A^\zeta\vu\>\,d\vx\>\\
&=0-\<\int_\Omega p A^\zeta\div\vu\,d\vx\>=0,
\end{align*}
using incompressibility. Therefore, $\(\cal B(\vu,\vu),A^\zeta\vu\)=0$ and in particular, $\(\cal
B(\vu,\vu),A^2\vu\)=0$.
\endofprf
\end{prf}

Since 2D incompressible homogeneous turbulence is widely believed
to be nearly isotropic far away from the influence of walls and
large-scale stirring forces, Theorem~\ref{triplet-vanishes} provides a
compelling argument that $(\cal B(\vu,\vu),A^2\vu)$ in
\Eq{NS-dim-palinstrophy} should be negligible. In fact, in
Fig.~\ref{tripletz1023} we show in a numerical simulation of fully
developed turbulence that the spatial contributions to the
triplet are tiny compared to the term
$\nu\|A^{3/2}\vu(t)||^2$ (twice the product of the viscosity $\nu$ and
the hyperpalinstrophy $P_2$).

Having proven the result in Theorem \ref{triplet-vanishes}, we can go further in studying
equations~(\ref{NS-dim-energy}), (\ref{NS-dim-enstrophy}), and (\ref{NS-dim-palinstrophy}).
Applying the Cauchy--Schwarz and Poincar\'{e} inequalities, we obtain
\begin{align*}
(\vf,\vu(t)) \leq \|\vf|| \|\vu(t)||, \quad k_0\|\vu(t)|| \leq \|A^{1/2}\vu(t)||,
\end{align*}
which leads to
\begin{equation*}
-\nu\|A^{1/2}\vu||^2 \geq -\nu k_0^2\|\vu||^2.
\end{equation*}
Thus, \Eq{NS-dim-energy} can be written as
\begin{equation}
\dfrac{d}{dt}\|\vu(t)||^2 \leq -2\nu k_0^2\|\vu(t)||^2 + 2\|\vf||\,\|\vu(t)||.
\label{NS-dim-energy-ineq}
\end{equation}
Simplifying the above inequality yields
\begin{equation}\label{E-NS}
\dfrac{d}{dt}\|\vu(t)|| \leq -\nu k_0^2\|\vu(t)|| + \|\vf||,
\end{equation}
which is a first-order differential inequality. If $\vf$ is constant in time, we
can solve the corresponding differential equation obtained by replacing the
inequality with an equality to obtain
\begin{equation*}
\|\vu(t)|| = e^{-\nu k_0^2t}\|\vu(0)|| + \(\dfrac{1-e^{-\nu k_0^2t}}{\nu
k_0^2}\)\|\vf||.
\end{equation*}
Applying the Gronwall inequality corresponding to \Eq{E-NS}, we thus find
\begin{equation}
\|\vu(t)|| \leq e^{-\nu k_0^2t}\|\vu(0)|| + \(\dfrac{1-e^{-\nu k_0^2t}}{\nu
k_0^2}\)\|\vf||. \label{E-NS-sol-2}
\end{equation}
Now, taking $\alpha \doteq e^{-\nu k_0^2t}$ and $\beta \doteq
\|\vf||/(\nu k_0^2)$, \Eq{E-NS-sol-2} can be expressed as
\begin{equation*}
\|\vu(t)||\leq\alpha\|\vu(0)||+(1-\alpha)\beta,
\end{equation*}
which is a segment connecting $\|\vu(0)||$ and $\beta$. On squaring both sides
and exploiting convexity, we obtain
\begin{equation*}
\|\vu(t)||^2 \leq \alpha\|\vu(0)||^2 + (1-\alpha)\beta^2.
\end{equation*}
We thus arrive at the following result:
\begin{equation}
\|\vu(t)||^2 \leq e^{-\nu k_0^2t}\|\vu(0)||^2 + (1-e^{-\nu k_0^2t})
\(\dfrac{\|\vf||}{\nu k_0^2}\)^2. \label{E-Gronwall-ineq}
\end{equation}
On introducing the Grashof number $G\doteq \|\vf||/(\nu^2 k_0^2)$, we
can write \Eq{E-Gronwall-ineq} as
\begin{equation}\label{E-Gronwall-Grashof-ineq}
\|\vu(t)||^2 \leq e^{-\nu k_0^2t}\|\vu(0)||^2 + (1-e^{-\nu k_0^2t})\nu^2 G^2.
\end{equation}
Applying the same argument to \Eq{NS-dim-enstrophy}, using \Eq{orthogonality}, results in a
similar estimate:
\begin{equation}
\|A^{1/2}\vu(t)||^2 \leq e^{-\nu k_0^2t}\|A^{1/2}\vu(0)||^2 + (1-e^{-\nu k_0^2t})\nu^2k_0^2G^2.
\label{Z-Gronwall-ineq}
\end{equation}
From \Eq{Z-Gronwall-ineq}, it can be observed that the closed ball
$\mathfrak{B}$ of radius $\nu k_0 G$ in the space $V$ is a bounded absorbing set
\cite{Dascaliuc2005}, and so weakly compact.\footnote{Every closed and
bounded convex set in a Hilbert space is compact in the weak topology.} If we
take~$S$ to be the solution operator for \Eq{NSFoias} defined by
\begin{equation*}
S(t)\vu_0=\vu(t), \qquad \vu_0=\vu(0)\in V,
\end{equation*}
where $\vu(t)$ is the unique solution \cite{Foias67} of \Eq{NSFoias}, then by the definition of the
absorbing set for the solution of a dynamical system, for any bounded set~\hbox{$\mathfrak{B'}\subset
V$}, there would be a time $t_0$ depending on a bound for $\mathfrak{B'}$ such that
\begin{equation*}
S(t)\mathfrak{B'} \subset
\mathfrak{B} ,\quad \forall t\geq t_0.
\end{equation*}
The global attractor $\cal A$ is then defined by
\begin{equation} \label{Global-Attractor}
\cal A= \Intersection_{t\geq0}{S(t)\mathfrak{B}},
\end{equation}
so $\cal A$ is the largest bounded, invariant set such that $S(t)\cal A = \cal
A$ for all $t\geq 0$. Taking into account in two dimensions
the existence of a global attractor
and a closed bounded absorbing set in $V \subset H$, an
immediate observation from \Eq{E-Gronwall-Grashof-ineq} and \Eq{Z-Gronwall-ineq}
shows that being on the attractor requires the following two conditions:
\begin{align}\label{bounds-on-attractor}
\|\vu|| &\leq \nu G, 		\\
\|A^{1/2}\vu|| &\leq \nu k_0 G.
\end{align}
The above observation leads to a suitable normalization for the energy and
enstrophy that we use later on for finding bounds in the $Z$--$E$ plane.
\begin{remark}
The above results assures us that on the attractor, both the energy and enstrophy are bounded.
\end{remark}

\section{Bounds on the projection of the global attractor onto the energy--enstrophy--palinstrophy
space}\label{bounds}

Having proven that $(\cal B(\vu,\vu), A^\zeta\vu)=0$ for $\zeta\in\R$, we can apply the same
argument applied in \cite{Emami18} to obtain upper bounds for the projection of the global
attractor on the $E$--$Z$ and $Z$--$P$ planes. However, that argument cannot be applied directly to find an
upper bound on the $E$--$P$ plane, though it is still possible to
obtain an upper bound using transitivity, as discussed in
Section~\ref{bounds-argument}.

Moreover, in view of Theorem~\ref{triplet-vanishes}, it is possible to extend this analysis to
generalized version of equations~(\ref{NS-dim-energy}), (\ref{NS-dim-enstrophy}), and
(\ref{NS-dim-palinstrophy}). In Section~\ref{bounds-argument}, we show that the method used to obtain an
upper-bound on the projection of the global attractor onto the $Z$--$E$ plane can be applied to any
sequence of consecutive polystrophies $E_\zeta, E_{\zeta+1},$ and $E_{\zeta+2}$. This important result
sheds light on self-similar behaviour of the global attractor.

\section{General polystrophies}\label{Genpoly}
\begin{definition}
For any solution $\vu$ of the 2D Navier--Stokes equation, we define
the $\zeta$th-order \emph{polystrophy}
\begin{equation*}
E_\zeta(t) = \half\|A^{\zeta/2}\vu(t)||^2,\qquad\zeta\in\R,
\end{equation*}
where $A\doteq-\lap$. Similarly, the $\zeta$th-order injection rate is defined as
\begin{equation*}
\epsilon_\zeta = \(\vf,A^\zeta\vu\).
\end{equation*}
\end{definition}
\noindent
In particular, $\epsilon_0:=\epsilon,\epsilon_1$, and $\epsilon_2$ are the
energy, enstrophy, and palinstrophy injection rates.
As discussed in~\cite{Emami18}, the Novikov theorem \cite{Novikov64}
guarantees that $\epsilon_\zeta \ge 0$ for every real $\zeta$.

Thus, for any $\zeta\in\R$ we obtain
\begin{align}
\(\frac{\partial \vu}{\partial t},A^\zeta\vu\)+\nu(A\vu,A^\zeta\vu)+(\cal B(\vu,\vu),A^\zeta\vu)
&=(\vf,A^\zeta\vu) \implies& \notag \\
\frac{1}{2}\dfrac{d}{dt}\|A^{\zeta/2}\vu||^2+ \nu\|A^{(\zeta+1)/2}\vu|| + (\cal B(\vu,\vu),A^\zeta\vu) &=
(\vf,A^\zeta\vu). \label{zeta-quadratic-dynamical-eq}
\end{align}
By Theorem~\ref{triplet-vanishes}, $(B(\vu,\vu),A^\zeta\vu)=0$ for any
$\zeta\in\R$; therefore
\begin{align}
\hf\dfrac{d}{dt}\|A^{\zeta/2}\vu||^2+ \nu\|A^{(\zeta+1)/2}\vu|| =
(\vf,A^\zeta\vu)=\epsilon_\zeta.
\label{zeta-quadratic-dynamic}
\end{align}
On applying the Poincar\'{e} inequality followed by a Gronwall
inequality, we find
\begin{align*}\label{zeta-quadratic-estimate}
\hf\dfrac{d}{dt}\|A^{\zeta/2}\vu||^2 &\leq \epsilon_\zeta - \nu k_0^2\|A^{\zeta/2}\vu||^2
\textoversign{Gronwall inequality}{\implies}\\
\|A^{\zeta/2}\vu(t)||^2 &\leq e^{-2\nu k_0^2t}\|A^{\zeta/2}\vu(0)||^2 +
\(\frac{1-e^{-2\nu k_0^2t}}{\nu k_0^2}\)\epsilon_\zeta.
\end{align*}
Thus, for every $\vu\in\cal A$, we would expect to have
\begin{equation}\label{energy-dynamical-wn}
\|A^{\zeta/2}\vu(t)||^2\leq \dfrac{\epsilon_\zeta}{\nu k_0^2}.
\end{equation}
\section{Bounds on the projection of the global attractor to the space of three
consecutive polystrophies}\label{bounds-argument}
Let $\vu(t)$ be a solution such that
$\vu(t)\neq \v0$ on some interval $(t_1,t_2]$. Then the function
$\chi(t)\doteq\frac{\|A^{\zeta/2}\vu||^2}{\|A^{\sigma/2}\vu||}$ for any $\zeta,\sigma\in\R$ such that
$\sigma<\zeta$ satisfies
\begin{align}\label{ki-1st-eq-wn}
\frac{d\chi}{dt}&=\dfrac{\dfrac{d\|A^{\zeta/2}\vu||^2}{dt}\|A^{\sigma/2}\vu||-\|A^{\zeta/2}\vu||^2
\dfrac{d\|A^{\sigma/2}\vu||}{dt}}{\|A^{\sigma/2}\vu||^2}\\\notag
&=\dfrac{2\(\epsilon_\zeta-\nu\|A^{(\zeta+1)/2}\vu||^2\)}{\|A^{\sigma/2}\vu||}-
\dfrac{\|A^{\zeta/2}\vu||^2\(\epsilon_\sigma-\nu\|A^{(\sigma+1)/2}\vu||^2\)}{\|A^{\sigma/2}\vu||^3}.
\end{align}
On defining,
\begin{equation*}
\lambda(t)\doteq\frac{\chi(t)}{\|A^{\sigma/2}\vu||}=
\frac{\|A^{\zeta/2}\vu||^2}{\|A^{\sigma/2}\vu||^2},
\end{equation*}
we see that \Eq{ki-1st-eq-wn} can be written as
\begin{equation}
\|A^{\sigma/2}\vu||\frac{d\chi}{dt}=-2\nu\|A^{(\zeta+1)/2}\vu||^2+2\epsilon_{\zeta}-\lambda
\epsilon_\sigma+\nu\lambda\|A^{(\sigma+1)/2}\vu||^2.\label{zeta-sigma}
\end{equation}
In the last equation we observe that if $\zeta=\sigma+1$, \Eq{zeta-sigma} can be simplified to
\begin{equation}
\|A^{\sigma/2}\vu||\frac{d\chi}{dt}=-2\nu\|A^{\sigma/2+1}\vu||^2+2\epsilon_\zeta-\lambda
\epsilon_\sigma+\nu\lambda\|A^{\zeta/2}\vu||^2.\label{zeta-sigma-simplified}
\end{equation}
Thus, on introducing $\vv=(A-\lambda)A^{\sigma/2}\vu-\frac{A^{\sigma/2}\vf}{2\nu}$, we obtain
\begin{align*}
\|\vv||^2&=\|(A-\lambda)A^{\sigma/2}\vu-\dfrac{A^{\sigma/2}\vf}{2\nu}||^2\\
&=\|A^{\sigma/2+1}\vu||^2-2\lambda\|A^{\zeta/2}\vu||^2-\dfrac{\epsilon_\zeta}{\nu}+
\dfrac{\lambda\epsilon_{\sigma}}{\nu}+\lambda^2\|A^{\sigma/2}\vu||^2+\dfrac{\|A^{\sigma/2}\vf||^2}{4\nu^2},
\end{align*}
so that
\begin{align*}
-2\nu\|\vv||^2&=-2\nu\|A^{\sigma/2+1}\vu||^2+4\nu\lambda\|A^{\zeta/2}\vu||^2+2\epsilon_{\zeta}
-2\lambda\epsilon_{\sigma}-2\nu\lambda^2\|A^{\sigma/2}\vu||^2\\
&\quad-\dfrac{\|A^{\sigma/2}\vf||^2}{2\nu}\\
&=\underbrace{\(2\epsilon_{\zeta}-2\nu\|A^{\sigma/2+1}\vu||^2\)-\lambda\epsilon_{\sigma}+
\nu\lambda\|A^{\zeta/2}\vu||^2}_{\displaystyle\|A^{\sigma/2}\vu||\dfrac{d\chi}{dt}}+\underbrace{
\nu\lambda\|A^{\zeta/2}\vu||^2}_{\displaystyle \nu\chi^2}\\
&\quad-\lambda\epsilon_{\sigma}-\dfrac{\|A^{\sigma/2}\vf||^2}{2\nu}.
\end{align*}
Thus
\begin{align*}
-\|A^{\sigma/2}\vu||\dfrac{d\chi}{dt}&=2\nu\|\vv||^2+\nu\chi^2-\lambda\epsilon_{\sigma}
-\dfrac{\|A^{\sigma/2}\vf||^2}{2\nu}.
\end{align*}
On introducing a real constant $\alpha$ whose value will be determined later, and considering the
fact that $\chi=\lambda\|A^{\sigma/2}\vu||$ we may write
\begin{align*}
\|A^{\sigma/2}\vu||\dfrac{d}{dt}(\alpha-\chi)&=2\nu\|\vv||^2+\nu(\alpha-\chi)^2-\nu\alpha^2
+2\nu\alpha\chi-\lambda\epsilon_{\sigma}-\dfrac{\|A^{\sigma/2}\vf||^2}{2\nu}\quad\\
&=2\nu\|\vv||^2+\nu(\alpha-\chi)^2-\nu\alpha^2+\(2\nu\alpha-\dfrac{\epsilon_{\sigma}}
{\|A^{\sigma/2}\vu||}\)\chi\\
&\quad-\dfrac{\|A^{\sigma/2}\vf||^2}{2\nu}.
\end{align*}
On defining $\beta=2\nu\alpha-\dfrac{\epsilon_{\sigma}}{\|A^{\sigma/2}\vu||}$,
the above result may be rewritten as
\begin{align*}
\|A^{\sigma/2}\vu||\dfrac{d}{dt}(\alpha-\chi)&=2\nu\|\vv||^2+\nu(\alpha-\chi)^2-\beta(\alpha-\chi)
+\alpha\beta-\nu\alpha^2-\dfrac{\|A^{\sigma/2}\vf||^2}{2\nu}.
\end{align*}
Thus, if $\alpha$ is such
that
\begin{align}\label{alpha-condition}
\beta=2\nu\alpha-\dfrac{\epsilon_\sigma}{\|A^{\sigma/2}\vu||}>0
\quad\text{and}\quad
\nu\alpha^2-\dfrac{\alpha\epsilon_{\sigma}}{\|A^{\sigma/2}\vu||}-\dfrac{\|A^{\sigma/2}\vf||^2}{2\nu}>0,
\end{align}
we can introduce
\begin{align*}
\phi\doteq 2\nu\|\vv||^2+\nu(\alpha-\chi)^2+\alpha\beta-\nu\alpha^2
-\dfrac{\|A^{\sigma/2}\vf||^2}{2\nu}>0
\end{align*}
to express the above first-order differential equation as
\begin{equation}
\|A^{\sigma/2}\vu||\dfrac{d}{dt}(\alpha-\chi)+\beta(\alpha-\chi)=\phi.
\end{equation}
The solution to this equation is given by
\begin{align*}
\alpha-\chi(t)&=(\alpha-\chi(t_0))\exp\(\int_{t_0}^{t}{-\dfrac{\beta}{\|A^{\sigma/2}\vu||}\,ds}\)\\
&+\int_{t_0}^{t}{\dfrac{\phi}{\|A^{\sigma/2}\vu||}\exp\(-\int_{\tau}^{t}
{\dfrac{\beta}{\|A^{\sigma/2}\vu||}\,ds}\)\,d\tau}.
\end{align*}
Taking $t_0\to -\infty$, and $t=0$, results in $\(\alpha-\chi(0)\)\geq 0$. Now taking $t_0=0$, and
$t\to \infty$, one finds that $\alpha-\chi(t)\geq 0$ for all $t\in(-\infty, \infty)$. Thus
\begin{equation}\label{wn-extended-result}
0\le \chi\leq\alpha\quad\implies\quad \|A^{\zeta/2}\vu||^2\leq\alpha\|A^{\sigma/2}\vu||.
\end{equation}
To get the above result we need to check conditions \Eq{alpha-condition}. Working on these
inequalities, one can show
\begin{align}\label{alpha-condition-2}
&\nu\alpha^2-\dfrac{\alpha\epsilon_\sigma}{\|A^{\sigma/2}\vu||}-\dfrac{\|A^{\sigma/2}\vf||^2}{2\nu}
=0\implies\\
&\alpha_{1,2}=\dfrac{\dfrac{\epsilon_\sigma}{\|A^{\sigma/2}\vu||}\pm\sqrt{\dfrac{\epsilon^2_\sigma}
{\|A^{\sigma/2}\vu||^2}+2\|A^{\sigma/2}\vf||^2}}{2\nu}\implies
\begin{cases}
\alpha_1 <0,\notag\\
\alpha_2 \geq \dfrac{\epsilon_\sigma}{\nu\|A^{\sigma/2}\vu||}.
\end{cases}
\end{align}
Therefore,
\begin{align*}
\nu\alpha^2-\dfrac{\epsilon_\sigma\alpha}{\|A^{\sigma/2}\vu||}-\dfrac{\|A^{\sigma/2}\vf||^2}{2\nu} >0
\iff \alpha >\alpha_2\ge \dfrac{\epsilon_\sigma}{\nu\|A^{\sigma/2}\vu||} \text{
  or } \alpha < \alpha_1 < 0, \notag
\end{align*}
which implies that
\begin{align*}
\beta=2\nu\alpha-\dfrac{\epsilon_\sigma}{\|A^{\sigma/2}\vu||}>0
\end{align*}
and, using \Eq{energy-dynamical-wn} (which implies that
$\epsilon_\sigma \ge 0$),
\begin{equation}
\alpha\geq\dfrac{\epsilon_\sigma}{\nu\|A^{\sigma/2}\vu||}\geq k_0\sqrt{\dfrac{\epsilon_\sigma}{\nu}}.
\label{alphaBound}
\end{equation}

If we set $\alpha=k_0\sqrt{\dfrac{\epsilon_\sigma}{\nu}}$, \Eq{wn-extended-result}
leads to a useful bound for polystrophies of order $\sigma$ and $\zeta=\sigma+1$:

\begin{equation}
\|A^{(\sigma+1)/2}\vu||^2 \leq k_0\sqrt{\dfrac{\epsilon_\sigma}{\nu}}\,\|A^{\sigma/2}\vu||.\label{gen-upper-bound-raw}
\end{equation}

Using \Eq{energy-dynamical-wn} and the Cauchy--Schwarz inequality,
we also find
\begin{equation}
k_0\sqrt{\nu\epsilon_\sigma}\leq\dfrac{\epsilon_\sigma}{\|A^{\sigma/2}\vu||}
=\fr{(\vf,A^\sigma\vu)}{\|A^{\sigma/2}\vu||}\le
\fr{\|A^{\sigma/2}\vf||\, \|A^{\sigma/2}\vu||}{\|A^{\sigma/2}\vu||}
=\|A^{\sigma/2}\vf||.
\label{C--S-NOV-Dynamical-combination}
\end{equation}
It is convenient to use the lower bound for $\|A^{\sigma/2}\vf||$ found in
\Eq{C--S-NOV-Dynamical-combination} to define a Grashof number for white-noise forcing:
\begin{equation*}
G=\dfrac{\|A^{\sigma/2}\vf||}{\nu^2 k_0^2}\ge \G\doteq\dfrac{\sqrt{\nu\epsilon_\sigma}}{\nu^2 k_0},
\end{equation*}
allowing us to express \Eq{gen-upper-bound-raw} more compactly:

\begin{theorem}\label{gen-upper-bound}
Let $\sigma\in\R$ and $\vu\in\cal A$ be a non-trivial solution of
\Eq{NSFoias} driven by a random forcing with $\sigma$-order polystrophy injection rate $\epsilon_\sigma$. On choosing $\G\doteq\sqrt{\nu\epsilon_\sigma}/(\nu^2 k_0)$, and introducing the
normalized velocity $\vv =\vu/(\G\nu)$,
\begin{equation*}
\|A^{(\sigma+1)/2}\vv||^2 \leq k_0^2\|A^{\sigma/2}\vv||.
\end{equation*}
\end{theorem}

The above result has the same form as found in \cite{Emami18} in the
$E$--$Z$ plane. We can go one step further and apply the above theorem twice to any three normalized
consecutive polystrophies:
\begin{equation*}
\|A^{(\sigma+2)/2}\vv||^2 \leq k_0^2\,\|A^{(\sigma+1)/2}\vv||\quad\text{and}\quad
\|A^{(\sigma+1)/2}\vv||^2 \leq k_0^2\,\|A^{\sigma/2}\vv||.
\end{equation*}
Therefore,
\begin{equation}\label{E--P-bound}
\|A^{\sigma/2+1}\vv||^4 \leq k_0^6\,\|A^{\sigma/2}\vv||.
\end{equation}
These generalized bounds are presented in Fig.~\ref{pzec1023}, along with a projection of the trajectory
of a numerical solution of 2D turbulence onto the energy--enstrophy--palinstrophy space.
\begin{remark}
It is important to observe that the inequality \Eq{E--P-bound} is quite special in the sense that it
cannot be directly derived by applying our analysis for $\sigma$ and $\sigma+2$ order polystrophies.
\end{remark}
\begin{remark}
In particular, for $\sigma=0$:
\begin{equation}
\frac{2P}{(\nu\G)^2}\leq k_0^2\sqrt{\frac{2Z}{(\nu \G)^2}}\leq
k_0^6\sqrt[4]{\frac{2E}{(\nu \G)^2}}.\label{PZEbounds}
\end{equation}
\end{remark}
\begin{observation}
It is important to observe that these relations are global on the additive subgroup of real
numbers that is homomorphic with the subgroup generated by $1$.
\end{observation}
\section{The inverse cascade and the necessity of hypoviscosity}\label{hypoviscosity}
In 3D turbulence, Kolmogorov argued that a uniform amount of energy
cascades from the forcing scales down to molecular scales, where it
gets dissipated \cite{Kolmogorov41}.
In 2D turbulence, Fj\oacc{}rtoft showed instead that a uniform amount of
enstrophy cascades down to the small scales, while a uniform amount of
energy cascades up to the large scales \cite{Fjortoft53}.
In numerical simulations of 2D turbulence in a bounded domain with doubly
periodic boundary conditions, this \emph{inverse cascade} causes the energy to
pile up at the large scales.
Although there is no pure realization of 2D turbulence in nature,
geophysicists believe that 2D turbulence can be roughly realized in high altitude atmospheric flow. It is
believed that the inverse energy cascade towards the large scales will be extracted from this realization
of 2D turbulence via interaction with gravity
waves. However, in pseudospectral simulations of
2D turbulence, where the boundary conditions are periodic,
there is no natural mechanism for extracting energy out
of these large scales. For this reason, numericists typically introduce
an artificial \emph{hypoviscosity} to take energy out of the flow
at large scales. Since there are various ways of implementing
hypoviscosity, it is crucial to understand how they affect the bounds
on the global attractor.

In this section we will analyze one of the simplest implementations of
hypoviscosity: a constant artificial friction $\gamma$
applied at all scales. The
governing equation for the incompressible 2D turbulence with random
forcing, \Eq{NSFoias}, becomes
\begin{equation}\label{NSfriction}
\frac{\partial \vu}{\partial t} + \nu A\vu + \cal B(\vu,\vu) + \gamma\vu = \vf.
\end{equation}
Consequently, the dynamical system equations \Eq{NS-dim-energy}, \Eq{NS-dim-enstrophy}, and
\Eq{NS-dim-palinstrophy} will change to
\begin{align*}
&\hf\ddt\|\vu(t)||^2+\nu\|A^\bhf\vu(t)||^2+(\cal B(\vu,\vu),\vu)+\gamma\|\vu||^2=(\vf,\vu(t)),\\
&\hf\ddt\|\Ahf\vu(t)||^2+\nu\|A\vu(t)||^2+(\cal B(\vu,\vu),A\vu)+\gamma\|\Ahf\vu||^2=(\vf,A\vu(t)),\\
&\hf\ddt\|A\vu(t)||^2+\nu\|A^{3/2}\vu(t)||^2+(\cal B(\vu,\vu),A^2\vu)+\gamma\|A\vu||^2=(\vf,A^2\vu(t)).
\end{align*}
Among the above equations, we are mainly interested in the energy and enstrophy equations. Taking into
account the bilinear identities, we find
\begin{align*}
\begin{cases}
(\cal B(\vu,\vu),\vu)=(\cal B(\vu,\vu),A\vu)=0,\\
\epsilon=(\vf,\vu),\quad\eta=(\vf,A\vu),
\end{cases}
\end{align*}
and therefore, we obtain the following system of differential equations:
\begin{align}
\begin{cases}
\ds\hf\ddt\|\vu||^2 + \nu\|A^\bhf\vu||^2 + \gamma\|\vu||^2 = \epsilon,\\[0.5cm]
\ds\hf\ddt\|\Ahf\vu||^2 + \nu\|A\vu||^2 + \gamma\|\Ahf\vu||^2 = \eta.
\end{cases}
\end{align}
Using the Poincar\'{e} inequalities
\begin{equation*}
k_0^2\|\vu||^2\leq\|\Ahf\vu||^2\qquad\text{and}\qquad k_0^2\|\Ahf\vu||^2\leq\|A\vu||^2,
\end{equation*}
we then obtain the following results for energy and enstrophy, respectively:
\begin{align}\label{energy-estimate-wn}
\hf\ddt\|\vu||^2 &\leq\epsilon-(\nu k_0^2+\gamma)\|\vu||^2 \textoversign{Gronwall inequality}{\implies}
\notag\\
\|\vu(t)||^2 &\leq e^{-2(\nu k_0^2+\gamma)t}\|\vu(0)||^2 +
\(\frac{1-e^{-2(\nu k_0^2+\gamma)t}}{\nu k_0^2+\gamma}\)\epsilon
\end{align}
and
\begin{align}\label{enstrophy-estimate-wn}
\hf\ddt\|\Ahf\vu||^2 &\leq\eta-(\nu k_0^2+\gamma)\|\Ahf\vu||^2 \textoversign{Gronwall inequality}{\implies}
\notag\\
\|\Ahf\vu(t)||^2 &\leq e^{-2(\nu k_0^2+\gamma)t}\|\Ahf\vu(0)||^2 +
\(\frac{1-e^{-2(\nu k_0^2+\gamma)t}}{\nu k_0^2+\gamma}\)\eta.
\end{align}
So for every $\vu\in\cal A$, we have
\begin{align}
\|\vu(t)||^2 &\leq\frac{\epsilon}{\nu k_0^2+\gamma},\label{u-on-the-attractor} \\
\|\Ahf\vu(t)||^2 &\leq\frac{\eta}{\nu k_0^2+\gamma}.\label{omega-on-the-attractor}
\end{align}
From \Eq{u-on-the-attractor}, we see that
$$
\sqrt{\nu\epsilon k_0^2 +
\epsilon\gamma} \leq\fr{\epsilon}{\|\vu||}=\fr{(\vf,\vu)}{\|\vu||}\le
\fr{\|\vf||\|\vu||}{\|\vu||}=\|\vf||.
$$

On introducing $\vf'=\vf-\gamma\vu$, which we call the \emph{reduced forcing}, the above differential
equations can be simplified to
\begin{align}
&\hf\ddt\|\vu(t)||^2+\nu\|A^\bhf\vu(t)||^2=(\vf',\vu(t)), \label{energy-friction-f'}   \\
&\hf\ddt\|\Ahf\vu(t)||^2+\nu\|A\vu(t)||^2=(\vf',A\vu(t)). \label{enstrophy-friction-f'}
\end{align}
On defining $\epsilon':=(\vf',\vu)=\e-\gamma \|\vu||^2$, and applying the argument in Section~\ref{bounds-argument} for
$\zeta=1,\sigma=0$, and considering $\vf'$ instead of $\vf$, the calculation of $\alpha$ and $\beta$
remains unchanged. The first inequality in~\Eq{alphaBound} then becomes
\begin{equation}
\alpha\geq\frac{\epsilon'}{\nu\|\vu||}.\label{alpha-epsilon-prime}
\end{equation}
Since $\epsilon\ge\epsilon'$, we can enforce \Eq{alpha-epsilon-prime} by
imposing
\begin{equation*}
\alpha\geq\frac{\epsilon}{\nu\|\vu||}\geq \frac{\sqrt{\nu\epsilon k_0^2 + \epsilon\gamma}}{\nu}.
\end{equation*}
Thus, on choosing $\alpha=\sqrt{(\nu\epsilon k_0^2 +
\epsilon\gamma)}/\nu$,~\Eq{wn-extended-result} becomes
\begin{equation}
\|\Ahf\vu||^2 \leq \frac{\sqrt{\nu\epsilon k_0^2 + \epsilon\gamma}}{\nu}\|\vu||.
\end{equation}
Therefore, the presence of a constant friction $\g$ changes the definition of
$\G$:
\begin{equation*}
G=\frac{\|\vf||}{\nu^2 k_0^2}\ge \G\doteq\frac{\sqrt{\nu\epsilon k_0^2 + \epsilon\gamma}}{\nu^2 k_0^2}.
\end{equation*}
This motivates the velocity normalization
$\vv(t)=\vu(t)/(\G\nu)$, so that
$$
\|\Ahf\vv||^2 \leq k_0^2\|\vv||.
$$
Thus, the presence of a constant friction (which we apply in our
numerical simulations to remove energy from the inverse cascade) 
leads to the same form for the bounds on the projection of
the global attractor onto the $E$--$Z$ plane.

\begin{remark}
The same argument with minor changes establishes that constant
forcing will not affect the form of the projection of the bounds of
the global attractor onto the $Z$--$P$ plane, as well as other
consecutive polystrophy pairs.
\end{remark}


\section{Pseudospectral numerical simulation}\label{pseudospectral}
To illustrate the theoretical results in this work,
we use a dealiased pseudospectral simulation to solve \Eq{NSfriction}
in a doubly periodic square box from zero initial conditions,
respecting the incompressibility condition $\div \vu = 0$.
For efficiency, we evolve the scalar \emph{vorticity} $\w=\zhat\dot\curl \vu$:
\begin{equation}
\frac{\partial \w }{\partial t}+(\hat{\bm{z}}\cross \grad \nabla^{-2}\w
\dot \grad )\w =-\gamma\vu +\nu_H\nabla^2\w +f,\label{NSvorticity}
\end{equation}
where $\hat \vz$ is the unit normal to the flow plane, $\nu_H$ is the
molecular viscosity, and $f$ is a white-noise random forcing.
As discussed in Section~\ref{hypoviscosity},
the friction term $\g$ prevents the energy injected
by $f$ from piling up at the largest scales.

Upon Fourier transforming, \Eq{NSvorticity} becomes
\begin{equation}
  \frac{\partial \w_{\vk}}{\partial t}=S_\vk
  -\nu_H k^2\w_{\vk} - \g \w_\vk +f_{\vk},\label{NShypoF}
\end{equation}
where $S_\vk=\sum_\vq\hat{\bm{z}}\vdot \vk\cross\vq\,
\w_{\vk-\vq}\w_{\vq}/q^2$ represents the advective convolution.
Although the theoretical treatment in Section~\ref{hypoviscosity} assumes $\g$ is constant, some
numericists replace it with a wavenumber-dependent hypoviscosity $H(k_L-k)\nu_L$, where~$H$ is the
Heaviside unit step function, which acts only at wavenumbers lower than $k_L$. This is to avoid
introducing unwanted dissipation in the inertial range, consistent with Kolmogorov's phenomenological
view of turbulence \cite{Kolmogorov41}. Most of our numerical results used a finite $k_L$ set
to the lowest forced wavenumber, but we verified numerically that replacing
the wavenumber-dependent friction with a constant friction $\gamma=\nu_L$ does not qualitatively
change the results, except that it steepens the inertial-range power
law of the energy spectrum well below the expected $k^{-3}$ power law obtained with the Heaviside cutoff.

We use $1023\times 1023$ dealiased Fourier modes to evolve \Eq{NShypoF} in a doubly periodic square box
of size $L=2\pi$; this is equivalent to $1534\times 1534$ spatial modes. In Fourier space, the turbulence
is driven by a white-noise forcing~$f_{\vk}$ limited to an annulus of mean radius $k_f=4$ and width
$\delta_f=1$. The parameters $\nu_H=4\times 10^{-5}$, $\nu_L=0.15$, and $k_L=3.5$ were chosen to give a
steady-state energy spectrum with a wide inertial range extending over roughly one decade of wavenumber
space. As described in~\cite{Emami18}, the Novikov theorem \cite{Novikov64} allows us to control the
average rate $\eta$ of enstrophy injection by the random forcing. We adopt the value $\eta=1$. An
adaptive time step was used to evolve the simulation from zero initial conditions to a fully developed
turbulent statistically stationary state. As is usual in numerical simulations of turbulence, we assume
that the ergodic theorem is sufficiently applicable so that ensemble averages may be approximated by
temporal averages.

The simulations were performed with the numerical code \texttt{DNS}
\url{https://github.com/dealias/dns/tree/master/2d} written by Professor John C. Bowman. The code,
written in C++, is comprised of a kernel called \texttt{TRIAD} that uses adaptive differential equation
solvers for ODEs and PDEs. The package contains a set of different numerical integrators and exploits the
\texttt{FFTW++} library \cite{fftwpp} for calculating hybrid dealiased convolutions \cite{Murasko24} under
Hermitian symmetry. Advanced computer memory management, such as implicit padding, memory alignment, and
dynamic moment averaging, allow \texttt{DNS} to attain its extreme performance. It uses the formulation
proposed by \citename{Basdevant83} to reduce the number of FFTs required for 2D (3D) incompressible
turbulence to four (eight). The reader who is interested in learning more about the \texttt{DNS} code is
referred to \url{https://github.com/dealias/dns/tree/master/2d}. Simplified 2D and 3D versions called
{\tt PROTODNS} have also been developed for educational purposes:
\url{https://github.com/dealias/dns/tree/master/protodns}.

We ran the simulations on a liquid-cooled Intel i9-12900K processor (5.2GHz, 8 performance cores) on an
ASUS ROG Strix Z690-F motherboard with 128GB of DDR5 memory (5GHz), using version  14.1.1 of the {\tt
GCC} compiler with the optimizations {\tt -Ofast -fomit-frame-pointer -fstrict-aliasing -ffast-math}. The
underlying FFTs were computed with version 3.3.10 of the adaptive {\tt FFTW} \cite{fftw,Frigo05} library
under the Fedora 40 operating system. Multithreading was implemented with the {\tt OpenMP} library.

The evolution of the trajectory of the numerical solution projected onto the
$E$--$Z$--$P$ space is shown in Fig.~\ref{pzez1023} along with the
three bounds predicted by~\Eq{PZEbounds} and the corresponding three
Poincar\'{e} inequalities. We observe that the solution lies well within
these bounds.

We verify in Figure~\ref{pzec1023} that allowing the friction to act at
all scales (by setting $k_L=\infty$) does not change the qualitative
nature of the bounds or the projected trajectory of the solution onto
the energy--enstrophy--palinstrophy space. As expected, the effect of
adding friction also to the small scales is to reduce the palinstrophy
and, to a lesser extent, the enstrophy.

\begin{figure}[htpb]
\includeWebFigure{pzez1023}{Projection of 2D forced--dissipative
turbulence with large-scale friction onto the $E$--$Z$--$P$ space.}
\end{figure}

\begin{figure}[htpb]
\includeWebFigure{pzec1023}{Projection of 2D forced--dissipative
turbulence with constant friction onto the $E$--$Z$--$P$ space.}
\end{figure}

In Fig.~\ref{vortz1023} we show a 2D colour density image of the
vorticity field of fully developed 2D turbulence, including
large-scale friction. A corresponding interactive 3D image of the
vorticity field using the same colour palette is shown in
Fig.~\ref{vort3dz1023}.

\begin{figure}[htpb]
\includeFigure{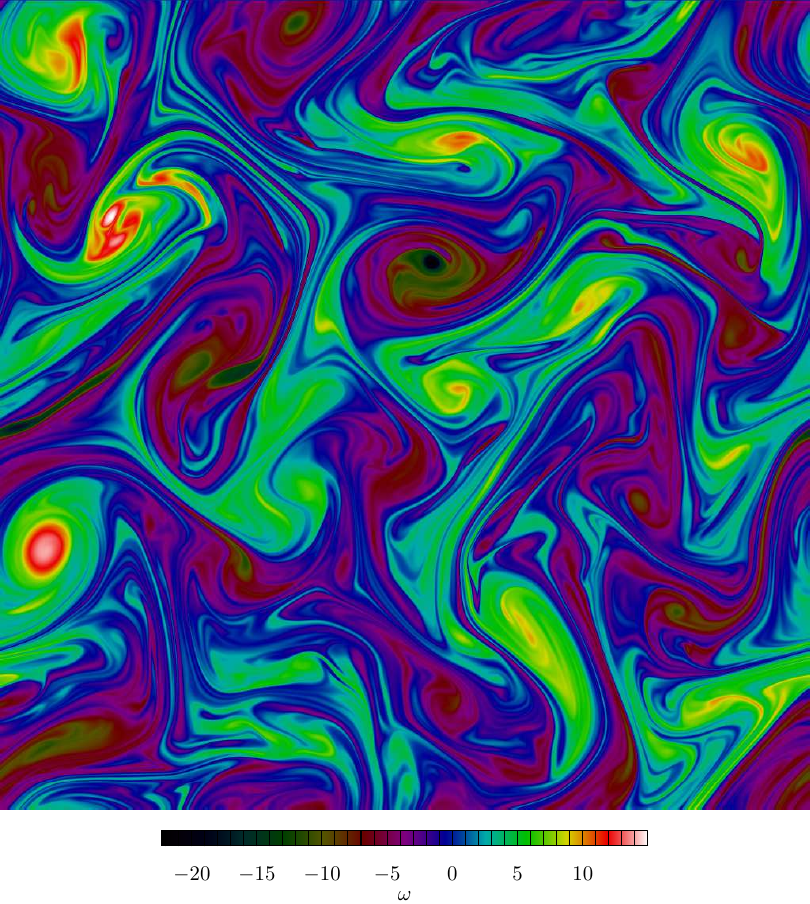}{Vorticity field at $t=446$ of 2D forced--dissipative
turbulence with large-scale friction.}
\end{figure}

\begin{figure}[htpb]
\includeWebFigure{vort3dz1023}{3D plot of the vorticity field at $t=446$ of 2D forced--dissipative
turbulence with large-scale friction.}
\end{figure}

In Fig.~\ref{ekz1023} we observe that the assumption of
isotropy underlying Theorem~\ref{triplet-vanishes} is well satisfied
for fully developed turbulence.

\begin{figure}[htpb]
  \includeWebFigure{ekz1023}{Isotropic energy spectrum averaged
    between $t=446$ and $t=566$ of 2D forced--dissipative turbulence
    with large-scale friction.}
\end{figure}

Consequently, we expect for fully developed turbulence that the
directions of the vectors $\B(\vu,\vu)$ and $A^2\vu$ are statistically
independent, with an expected alignment angle
$$
\<\theta\>\doteq\cos\inv\(\frac{\<(\vu\dot\grad)\vu\dot A^2\vu\>}{\sqrt{\<|(\vu\dot\grad)\vu|^2\>\<|A^2\vu|^2\>}}\)
$$
approaching $90^\degrees$.

We observe in Fig.~\ref{angle1z1023init} that the
average of $\theta$ over the first $0.1$ time units of the evolution
is spatially smooth. As the turbulence evolves, however, this smooth field
becomes spatially randomized, so that for fully developed turbulence
one obtains
the randomized values of $\<\theta\>$ seen in Fig.~\ref{angle1z1023},
when averaged over the same short time interval.
However, as one averages over longer time intervals, the spatial
fluctuations in $\<\theta\>$ are reduced and begin to approach the
theoretically predicted value of $90^\degrees$, as seen in Fig.~\ref{anglez1023}.

\begin{figure}[htpb]
  \includeWebFigure{angle1z1023init}{Average from $t=0$ to $t=0.1$ of the alignment
    angle $\theta$ between $\B(\vu,\vu)$ and $A^2\vu$ in the transient phase.}
\end{figure}

\begin{figure}[htpb]
  \includeWebFigure{angle1z1023}{Average
    from $t=446$ to $446.1$ of the alignment
    angle $\theta$ between $\B(\vu,\vu)$ and $A^2\vu$
    in fully developed turbulence.}
\end{figure}

\begin{figure}[htpb]
  \includeWebFigure{anglez1023}{Average
    from $t=446$ to $t=566$ of the alignment
    angle $\theta$ between $\B(\vu,\vu)$ and $A^2\vu$ in fully developed turbulence.}
\end{figure}

The evident lack of alignment between $\B(\vu,\vu)$ and $A^2\vu$
is responsible for the smallness observed in Fig.~\ref{tripletz1023}
of the triplet $(\cal B(\vu,\vu),A^2\vu)$, relative to the term $2\nu
P_2$ in \Eq{NS-dim-palinstrophy}.
The triplet $(\cal B(\vu,\vu),A^2\vu)$ was computed as a ternary
convolution, using the advanced hybrid dealiasing framework
available in version 3.02 of the \texttt{FFTW++} library
\cite{Murasko24,fftwpp}.

\begin{figure}[htpb]
  \includeWebFigure{tripletz1023}{Average from $t=446$ to $t=566$ of the
    normalized triplet contributions $\langle(\vu\dot\grad)\vu\dot A^2\vu\rangle/(2\nu P_2)$ in fully developed turbulence.}
\end{figure}

If the large-scale energy-extracting hypoviscosity is omitted from
\Eq{NShypoF}, the resulting pile-up of energy leads to the
formation of the spatial dipole vortex pair seen in
Figs.~\ref{vortz1023pure} and~\ref{vort3dz1023pure}.
These large-scale vortices allow the flow to interact with the
boundaries of the arbitrarily imposed square domain (subject to
periodic boundary conditions).
The interaction of the dipole vortices with the square boundaries results
in the highly anisotropic energy spectrum seen in
Fig.~\ref{ekz1023pure}. In the absence of friction,
the lack of isotropy prevents one from applying
Theorem~\ref{triplet-vanishes} to the long-term dynamics of
turbulence in a square domain with periodic boundary conditions.
Not surprisingly, we see in
Fig.~\ref{pzez1023pure} that the projection of the trajectory of the
numerical solution onto the energy--enstrophy--palinstrophy space
is no longer restricted to lie within the bounds given by \Eq{PZEbounds}.

\begin{figure}[htpb]
  \includeFigure{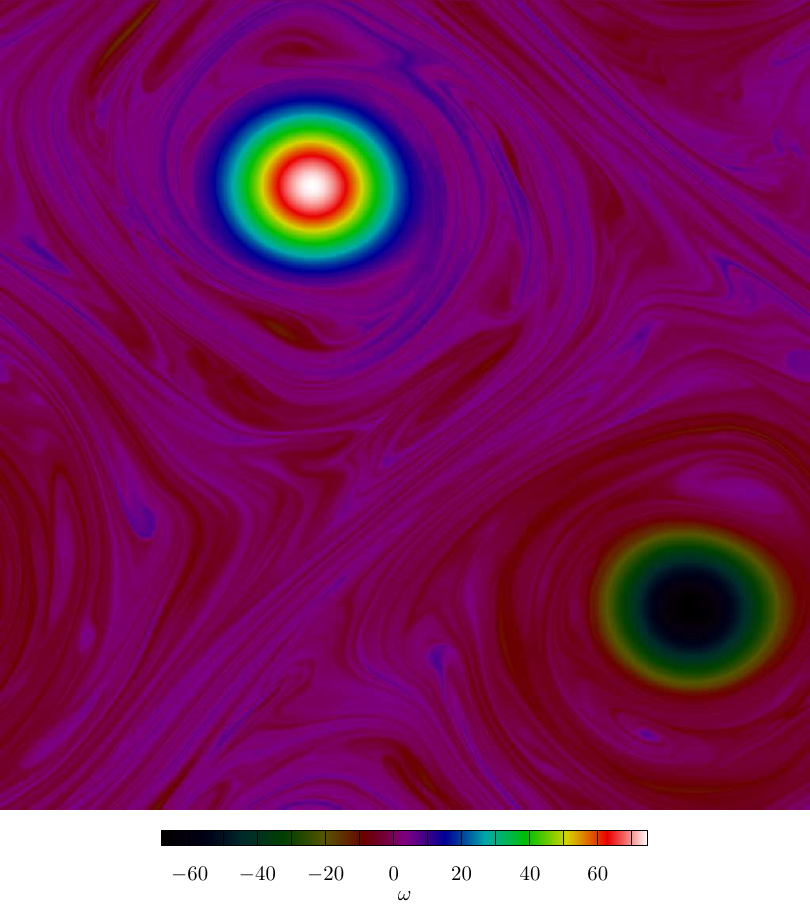}{Dipole vortices at $t=591$ of pure 2D
forced--dissipative turbulence.}
\end{figure}

\begin{figure}[htpb]
\includeWebFigure{vort3dz1023pure}{3D plot of the dipole vortices at $t=591$ of pure 2D
forced--dissipative turbulence.}
\end{figure}

\begin{figure}[htpb]
  \includeWebFigure{ekz1023pure}{Anisotropic energy spectrum averaged
    between $t=1000$ and $t=1120$ of pure 2D forced--dissipative turbulence.}
\end{figure}

\begin{figure}[htpb]
\includeWebFigure{pzez1023pure}{Projection of pure 2D forced--dissipative
turbulence onto the $E$--$Z$--$P$ space.}
\end{figure}

\section{Discussion}
In previous work \cite{Emami18}, we showed that the upper bound in
the $E$--$Z$ plane obtained by \citename{Dascaliuc2005} for
constant forcing also works for white-noise random forcing under the
proper normalization. Unfortunately, the bounds that
\citename{Dascaliuc2010} obtained in the $Z$--$P$ plane vastly
overestimate actual values from numerical simulations.
In this work,
we show this bound can be greatly tightened by exploiting isotropy.
The key difficulty in refining these bounds is the lack of a good
estimate for the triplet term $(\B(\vu,\vu),A^2\vu)$.
However, for statistically isotropic turbulence with arbitrary forcing,
we have shown for any real $\zeta$ that the expected value of the
triplet $(\B(\vu,\vu),A^\zeta\vu)$ is zero.
Moreover, if the turbulence is driven by
a white-noise random forcing, one can make use of the Novikov theorem
\cite{Novikov64} to go further and obtain the same
form of the upper bounds for $P$ vs.\ $Z$ as we derived for $Z$ vs.\ $E$
in \cite{Emami18}. In addition, we saw that enstrophy can be used as
an intermediary to provide new $P$ vs.\ $E$ bounds, using transitivity.
Combining these results with the corresponding Poincar\'{e} inequalities
leads to six bounds in the $E$--$Z$--$P$ space.
Likewise, the general inequality we derived for consecutive polystrophies
can be used to obtain relations between any polystrophy pair.

Furthermore, we showed that a slightly modified normalization can be
used to extend all of these results to account for the large-scale
friction (hypoviscosity) that is normally used by numericists to
remove energy at the large scales and thereby ensure isotropy.
We showed that realistic numerical simulations of forced--dissipative
turbulence respects these bounds. We also confirmed that
$\B(\vu,\vu)$ and $A^2\vu$ are statistically independent and showed that
$(\B(\vu,\vu),A^2\vu)$ is negligible compared to the dissipative term of the
governing equation.

The analytical bounds we have derived for 2D turbulence driven by
random forcing provides a mathematical framework to evaluate various heuristic
turbulent subgrid (and supergrid) models by characterizing the
behaviour of the global attractor under these models.

\appendix

\section{Inequalities}\label{app-C}
Among many inequalities in the field of functional analysis, we now list those that are used in this
work. The reader who is interested in studying their proofs and related inequalities, especially
\In{Sobolev inequalities}, can refer to any standard functional analysis resource.

In this appendix we use the following notation:
\begin{itemize}
\item $\Omega $: the domain of definition of functions;
\item $\vu$: $\vu(\vx), \quad \vx \in \Omega\subset \mathbb{R}^3$, or $\Omega\subset \mathbb{R}^3$;
\item $\vu_{\vk}$: the discrete Fourier transform of $\vu,\quad \vk \in
\mathbb{Z}^3\backslash\{\bm{0}\}$, or $\vk \in \mathbb{Z}^2\backslash\{\bm{0}\}$;
\item $A=-\lap$: a self-adjoint linear operator. 
\end{itemize}
For the sake of simplicity, we are going to ignore $\Omega$ in our
notation; for example, we would write $H^2$ instead of $H^2(\Omega)$. 

\section{Agmon inequalities}
Let $\Omega\subset \R^3$ and $\vu\in H^2(\Omega)\cap H_{0}^1(\Omega)$. Then the 3D Agmon inequalities state
that there exists a constant $C_A$ such that
\begin{align*}
\|\vu||_{L^\infty(\Omega)}&\leq C_A\|\vu||_{H^{1}(\Omega)}^{1/2}\|A\vu||_{H^2(\Omega)}^{1/2},\\
\|\vu||_{L^\infty(\Omega)}&\leq C_A\|\vu||_{L^2(\Omega)}^{1/4}\|A\vu||_{H^2(\Omega)}^{3/4}.
\end{align*}
When $\Omega\subset\mathbb{R}^2$ and $\vu\in H^2(\Omega)\cap
H_{0}^{1}(\Omega)$, the 2D Agmon inequality states that there exists a constant $C_A$ such that
\begin{equation}
\|\vu||_{L^\infty(\Omega)}\leq C_A\|\vu||^{1/2}\|A\vu||^{1/2}.\label{Agmon2D}
\end{equation}
It must be noted here that $C_A$ is a universal constant that
depends only on the space $\Omega$, so in order to make use of this inequality for numerical simulations,
having an explicit value for $C_A$ is crucial. In the next section we
are going to evaluate this constant for the two-dimensional periodic case.

\section{Universal constant of the Agmon inequality in a 2D periodic domain}
The main reason for finding an explicit value for the constant in the Agmon inequality is to obtain some
useful bounds for our direct numerical simulation of 2D turbulence. As the numerical simulation is done
on a lattice, we focus on the discrete form of the Agmon inequality,
where the integrals are replaced by summations. Because our simulations are performed in Fourier space, we make use
of the Fourier representation of the Agmon inequality.

We consider continuous functions on a bounded domain. The numerical simulations are done on a lattice on
this domain, where suprema reduce to maxima. That is,
\begin{equation*}
\|\vu||_{L^\infty(\Omega)} =\max_{\Omega}|\vu|.
\end{equation*}
It is worth mentioning that \In{pseudospectral} numerical simulations of turbulence in a periodic
domain use the $N$-point discrete Fourier transform
\begin{equation*}
\hat\vu_k=\dfrac{1}{\sqrt{N}}\sum_{j=0}^{N-1} \vu_j e^{-\fr{2\pi ijk}{N}},
\end{equation*}
which has the inverse
\begin{equation*}
\vu_j=\dfrac{1}{\sqrt{N}}\sum_{k=0}^{N-1} \hat\vu_k e^{\fr{2\pi ijk}{N}}.
\end{equation*}
The orthogonality relationship underlying this transform pair is elucidated on substituting $z=e^{2\pi
i(j+\bar j)/N}$:
\begin{equation*}
\sum_{k=0}^{N-1} e^{2\pi i(j+\bar j) k/N}=
\sum_{k=0}^{N-1} z^k=
\begin{cases}
N &\text{if}\quad j+\bar{j}=mN,\quad m\in\Z,  \\
\frac{1-z^N}{1-z}=0 &\text{otherwise}.
\end{cases}
\end{equation*}
Taking into account the above definitions, one can easily obtain the convolution theorem for the product
of two functions $f\text{ and }g$:
\begin{equation*}
\dfrac{1}{\sqrt{N}}\sum_{j=0}^{N-1} f_jg_j e^{-\fr{2\pi ikj}{N}}=
\dfrac{1}{N\sqrt{N}}\sum_{j=0}^{N-1}e^{-\fr{2\pi ikj}{N}}\sum_{p=0}^{N-1}\hat{f_p}
e^{-\fr{2\pi ipj}{N}}\sum_{q=0}^{N-1}\hat{g_q} e^{-\fr{2\pi iqj}{N}}.
\end{equation*}
Using the orthogonality condition
\begin{equation*}
\sum_{j=0}^{N-1}e^{-\fr{2\pi i(p+q-k)j}{N}}=N\delta_{p+q-k,0},
\end{equation*}
yields
\begin{equation*}
\dfrac{1}{\sqrt{N}}\sum_{j=0}^{N-1} f_jg_j e^{-\fr{2\pi
ikj}{N}}=\dfrac{1}{\sqrt{N}}\sum_{j=0}^{N-1}\hat{f_p}\hat{g}_{k-p}.
\end{equation*}
So, taking $k=0$ and $g=f$, we obtain the \In{Parseval identity}
\begin{equation}\label{Parseval}
\sum_{j=0}^{N-1} f_j^2 = \sum_{p=0}^{N-1}\hat{f}_p\hat{f}_{-p} = \sum_{p=0}^{N-1}\|\hat{f}_p|^2.
\end{equation}
Now using the fact that
$\(\max_{\Omega}|\vu|\)^2  \leq \sum_{\vx}{|\vu|^2}$ and $\sqrt 2
\le\sqrt{1+k^2}$ for $\vk\in\Z^2\backslash\{\bm{0}\}$, and applying the
Parseval identity \Eq{Parseval} and \In{H\"older inequality}, we find
\begin{align}\label{Fourier}
\(\max_{\Omega}|\vu|\)^2 \leq \sum_{\vx}{|\vu|^2} &=\sum_{\vk}{|\vu_{\vk}|^2} \leq
\fr{1}{\sqrt2}\sum_{\vk}{|\vu_{\vk}|^2\sqrt{1+k^2}}\ \textoversign {H\"older Ineq.}{\leq}
\notag\\ 
&\fr{1}{\sqrt2}\(\sum_{\vk}{|\vu_{\vk}|^2}\)^{1/2}\(\sum_{\vk}{|\vu_{\vk}|^2(1+k^2})\)^{1/2}.
\end{align}
Consequently,
\begin{align}\label{Agmon-Fourier}
\max_{\Omega}|\vu| \leq
\dfrac{1}{\sqrt[4]2}\(\sum_{\vk}{|\vu_{\vk}|^2}\)^{1/4}\(\sum_{\vk}{|\vu_{\vk}|^2(1+k^2})\)^{1/4}.
\end{align}
Comparing to \Eq{Agmon2D}, we thus see that
\begin{equation*}
C_A=\dfrac{1}{\sqrt[4]2}.
\end{equation*}

\section*{Acknowledgments}
Financial support for this work was provided by Discovery Grant
RES0043585 from the Natural Sciences and Engineering Research Council
of Canada.

\clearpage
\printbibliography[heading=bibintoc]
\end{document}